\documentstyle[prc,preprint,eqsecnum,aps]{revtex}
\tightenlines
\newcommand{\eeq}{\end{equation}}
\newcommand{\beq}{\begin{equation}}
\newcommand{\beqa}{\begin{eqnarray}} 
\newcommand{\eeqa}{\end{eqnarray}} 

\newcommand{\eqn}[1]{\nolinebreak{Eq.\ (\ref{#1})}}    
\input{psfig}    
\begin{document}
\title{Collision Broadening of the \boldmath{$\phi$} Meson in Baryon 
Rich Hadronic Matter}
\author{Wade Smith\footnote{Permanent Address:  Department 
of Physics, SUNY-Stony Brook, Stony Brook, NY 11794-3800\\
Electronic Address: smith@skipper.physics.sunysb.edu}
and Kevin L. Haglin\footnote{Permanent Address: Department of Physics,
Grinnell College, P.O. Box 805, Grinnell, IA 50112\\
Electronic Address: haglin@ac.grin.edu}}
\address{Department of Physics\\ Lawrence University\\ Appleton, WI 54912}
\date{\today}
\maketitle
\begin{abstract}
Phi meson-baryon cross sections, estimated within a one-boson-exchange 
model, serve as input for a calculation of the collision rates in hot 
hadronic matter.  We find that the width of the $\phi$ meson 
is modified through collisions with baryons by 
1--10 MeV at 160 MeV temperature depending on the baryon
fugacity.  Thermalization of the $\phi$  in 
high energy heavy ion collisions is discussed. 
\end{abstract}


\section{Introduction}

One of the premier goals of relativistic heavy-ion physics
is to identify and describe the quark gluon plasma (QGP).  There has been much 
work directed towards quantifying the properties of QGP  
\onlinecite{asakawa93,lee75,wilczed94,muller92}.
Understanding the physics in absence of plasma is necessary preparation.
Vector mesons provide a useful probe in 
measuring the properties of the
medium, and in detecting the predicted crossover to the QGP.
The $\phi$ is especially promising because it decays  both to kaon pairs
$(\sim 83\%)$
and, more rarely, to lepton pairs, both of which 
are readily detectable in high energy
nuclear experiments\cite{pdb}.  
It is  clear, however, that the in-medium properties of the $\phi$ are
most directly detected through its leptonic decays\cite{siebert95}.
Reconstructing the $\phi$ mass through the products of the dominant 
branching ratio, $\phi \rightarrow KK$, does not suffice
since detectable kaons emerge at freezeout;
kaons from $\phi$ decays in the medium are unlikely to escape
without reacting further, thus destroying any useful information
possessed about $\phi$.  
The decay of $\phi$ to leptons, however, is
sufficiently rare that previous experiments have not been able
to fully exploit the $\phi$ as an in-medium probe.  The greater
luminosities provided by the next generation of colliders should
render the $\phi$ an invaluable tool in quantifying in medium properties.

The $\phi$ makes a nice probe since it is not masked behind other
resonances in mass spectra.
Whence, much work has been done predicting how the properties of the
$\phi$ change in the medium.  Due to partial restoration of chiral symmetry
\cite{brown96},
the mass of the $\phi$ may be altered.  Such studies have been carried
out intensively using a variety of methods.  The  $\phi$ mass has been studied
through the use of  
effective chiral Lagrangian approaches\cite{kuwabara95,song96,klingl97}
notably through
the use of a full SU(3) chiral Lagrangian with SU(3)$_v$-breaking
effects\cite{bhatt97},
using QCD sum rules\cite{asakawa94,martinovic90},  the
Nambu-Jona-Lasinio model\cite{blaizot91}, and with others.   
These approaches tend to report a modest drop in the mass of
the $\phi$ with increasing temperature.  

The topic of this paper, the width of the $\phi$, has also been studied 
previously.  These studies, however, have focused on effects resulting
from a $\phi$ mass dropping scenario
due to chiral symmetry
restoration, or similarly, through kaon mass dropping effects
\cite{bhatt97,martinovic90,shuryak92,lissauer91,panda93,ko94,bhalerao97}.
This work presents a systematic study of $\phi$ dynamics by calculating
the change in the width of the $\phi$ through scattering with other particles
in the medium.  We include a rather complete set of important 
$\phi$-meson scattering reactions which 
contribute to the collision broadening.
In contrast with the earlier works listed, we do not compute
a modification in the {\em decay} width of the $\phi$, rather, we 
calculate a modification in the full width due to vigorous
scattering.
The bare mass for the $\phi$ is taken throughout, and  dropping
mass effects are not considered here.  A similar collision broadening 
study has been carried out for the $\rho$ meson in which mass dropping
was considered\cite{klh96}.

\section{The Model}

We examine conditions typical of heavy-ion collisions at the
Alternating Gradient Synchrotron (AGS) up to the Super Proton 
Synchrotron (SPS).  Collisions of these types reach  temperatures 
ranging from 120-160 MeV or higher. Recently  
observed pion-proton ratios suggest interpretation of a baryon
chemical potential in the neighborhood  of 400 MeV\cite{rapp97}.  We 
take this as one possible value in our study.    

The calculation discussed arises from an effective kinetic theory,
based on a fireball model.
We approximate  distributions of the various particles 
as Boltzmann.

\subsection{Particle Density}

The probability that a $\phi$ meson will collide with another particle 
depends upon both 
the reaction cross section  and
the population density of the partner.
We assume  Boltzmann distributions for simplicity.  The density 
is a function of temperature,  and for  particle $i$ is 
given by
\beq
\frac{dn_i}{d^3x}
=
d_i
\int^{\infty}_0 \frac{d^3p_i}{(2\pi)^3}e^{-\beta E}, 
\label{density1}
\eeq
where $\beta$ is inverse temperature and $d_i$ is the degeneracy.
The degeneracy is given by, 
\beq
d_i =
(2I_i+1)(2S_i+1),
\eeq
where $I_i$ is the isospin of the particle,  and $S_i$ is the spin.  
For the $\phi$,  \eqn{density1} can be simplified to read
\beq
\frac{dn_{\phi}}{d^3x}
=
\frac{3m^2_{\phi}K_2(\frac{m_{\phi}}{T})T}{2\pi^2}, 
\label{phidensity}
\eeq
where $K_2$ is a second order modified Bessel function.

\subsection{The Reaction Cross Section}

A fundamental ingredient in the calculation is the reaction cross
section.
Here we use boson exchange to model the interactions.  Starting with
the Lagrangians,
\beqa
{\cal L}_{NN\pi} &=& g_{NN\pi}\bar\psi\gamma_5 \vec\tau\psi \cdot \vec\pi \\
{\cal L}_{NN\rho} &=& g_{NN\rho} \bar\psi \gamma_\mu \vec\tau \psi \cdot 
\vec\rho^{\,\mu} +
\frac{f_{NN\rho}}{4m_N}\bar\psi\sigma_{\mu\nu}\psi\vec\rho^{\,\mu\nu} \\
{\cal L}_{N\Delta\pi} &=& \frac{f_{N\Delta\pi}}{m_\pi} \bar\psi \vec T 
\psi_\mu \cdot \partial^\mu \vec\pi \\
{\cal L}_{\phi\rho\pi} &=& \frac{g_{\phi\rho\pi}}{m_\pi} 
\epsilon_{\mu\nu\alpha\beta}
\partial^\mu \rho^\nu \partial^\alpha \phi^\beta \pi \\
{\cal L}_{KK\phi} &=&  g_{\phi KK} \phi^\mu (\partial_\mu \vec K
\times \vec K)_{(3)},
\eeqa
we generate a set of tree-level graphs to compute the squared
amplitude.  The cross section is then computed with 
\beq
\sigma=\frac1F I(s),
\label{byck2.2}
\eeq
where
\beq
F=2\lambda^{1/2}(s,m_a^2,m_b^2)(2\pi)^2
\label{byck2.3}
\eeq
is the flux factor\footnote{Here $\lambda$ is the standard kinematical
function\cite{byckl} defined 
by $\lambda(a,b,c) = a^2+b^2+c^2-2ab-2bc-2ca$, and sometimes called 
the triangle function since
$\frac14\sqrt{-\lambda(a^2,b^2,c^2)}$ gives the area of a triangle with sides 
$a,b,c$.} 
and
\beq
I_2(s)=\int \frac{d^3p_1}{2E_1}\frac{d^3p_2}{2E_2}\ \delta^4
\left(
p_a + p_b - p_1 - p_2
\right)
|{\cal M}|^2
\label{byck2.4}
\eeq
contains the integration over phase space.

The model thus far has not taken into account the fact that the particles 
involved are extended objects.  A multiplicative 
factor must be inserted to suppress
high momentum transfers.
Each vertex in a t-channel 
Feynman diagram requires a multiplicative factor attached to the 
scattering amplitude of the form \cite{klh95},
\beq
h=\frac{(\Lambda^2-\zeta)^2}{(\Lambda^2-t)^2}, 
\label{formf}
\eeq
where $\zeta$ is the squared mass of the exchanged virtual meson.  If the
exchanged meson
can go on shell, $\zeta$ is given by\cite{byckl}
\beq
\zeta=t_{max}=m_a^2+m_1^2-\frac{1}{2s}
\left[
(s+m_a^2-m_b^2)(s+m_1^2-m_2^2)-\lambda^{1/2}(s,m_a^2,m_b^2)
\lambda^{1/2}(s,m_1^2,m_2^2)
\right],
\eeq
to assure that $h$ is bounded from above by unity.

\subsection{The Thermally Averaged Cross Section}

We  consider reactions that occur in the central region
of relativistic heavy ion collisions.  The distributions of the 
particles composing
this hot hadronic matter  
will be assumed approximately thermal.
The cross section, $\sigma(\sqrt{s})$, is a function of a fixed total
energy of the initial state.  
We  must 
sum over $\sigma(\sqrt{s})$
for each  value of $\sqrt{s}$ that may occur in the
thermal distribution, weighting each cross section's contribution to the sum
according to the probability of having that particular configuration.
This is denoted as the thermally averaged cross section. 
\beq
\overline\sigma = 
\frac{1}{4m_N^2m_{\phi}^2K_2(\frac{m_N}{T})K_2(\frac{m_{\phi}}{T})}
\int^\infty_{\frac{m_N+m_{\phi}}{T}}dz_1K_1(z)
\lambda(s,m_N^2, m_{\phi}^2)\sigma_{\phi N}(\sqrt{s}),
\label{sigbar}
\eeq
where $s=z^2T^2$, and $z$ is a dimensionless energy variable.

\subsection{The Collision Rate, $\overline{\Gamma}$}

The decay rate, 
$\Gamma$, is the probability per unit time that a particle will  decay.
Using natural units,  where $\hbar=c=1$ one has simply 
\beq
\Gamma = \frac{1}{\tau}.
\eeq
For the $\phi$ meson, 
$\tau_\phi = 15 \times 10^{-23} \mbox{\rm s},
\ \Gamma_\phi = 4.43 \mbox{\rm MeV}.$
These  are the free space values, indicating the mean lifetime
and the decay rate of the $\phi$ with no external forces acting upon it.  

When a
particle is situated within the spacetime evolution  
of a heavy ion collision, for example, 
these values could change due to the extreme temperatures
and densities.   There is a significant 
probability that the $\phi$ will scatter with other particles in the medium.  
The scattering randomly kicks the $\phi$ meson's quantum mechanical
phase, effectively giving the distribution a broadened appearance.  
In equilibrium, however, there are as many $\phi$-producing
reactions as there are $\phi$-reducing ones.
Thus, the number does not change, only the distribution does.
  
The collision rate, $\overline\Gamma$, 
is now approximated by, 
\begin{equation}
\overline\Gamma = n\,\overline\sigma
\label{coll}
\end{equation}
where $n$ is the particle density given in \eqn{density1}, and 
$\overline{\sigma}$ is given by \eqn{sigbar}.
Note that the relative velocity is  included within 
$\overline{\sigma}$.
The results for several reactions are
included below.

\subsection{The Mean Free Path}

Another useful measure of collision rates is the mean free path, $\lambda$,  
which gives a sense of relaxation toward thermalization.  
It is not the intent of this work to provide a rigorous exploration
of the rate equations.  Rather, an estimate of the
mean free path shall be used as a tool to interpret our findings.  
We use, 
\beq
\lambda = \tau\,\overline{v} = \frac{\overline{v}}{\Gamma},
\label{mfp}
\eeq
where $\overline{v}$ is the average velocity of the $\phi$ in the medium. 

The extent in configuration space of the hadronic matter formed in 
heavy-ion collisions is not precisely known, but a reasonable 
first estimate comes from a Bjorken model.  Upper
limits of 10--15 fm are typical.  For a given distribution of $\phi$'s 
there is finite chance for decay inside, but many will decay outside.  We 
explore consequences of those rare $\phi$'s that  decay inside the medium.

\section{Results}

\begin{figure}
\centerline{\psfig{figure=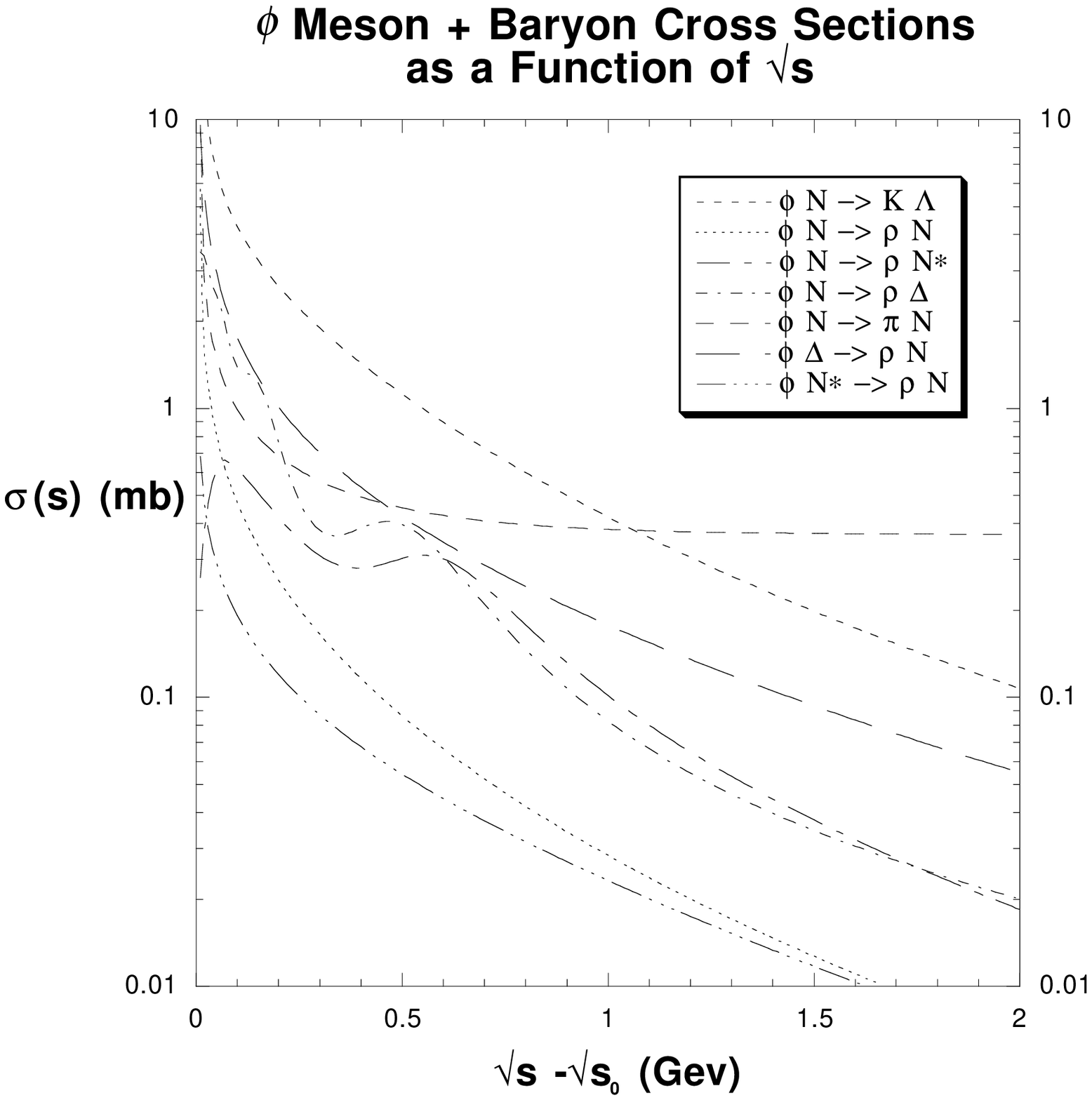}
	    }
\caption{Cross sections for several $\phi$ + baryon reactions.}
\label{xsects}
\end{figure}

The contributions of seven reactions were computed.
The cross section for each individual reaction is plotted in 
Fig.\ \ref{xsects}.
Of the reactions shown, the cross section for the reaction
$\phi {\rm N} \rightarrow {\rm K} \Lambda$ with kaon exchange is larger than
the others 
in the region below $\sqrt{s_0}+1.0$ GeV, where $\sqrt{s_0}$
is the reaction's threshold energy.  This energy region is most important 
since reactions within this domain will be the most common in the nuclear 
medium.  The relatively large cross section for 
this reaction is due primarily to the large couplings.
In the absence of experimental data  for 
the nucleon-kaon-$\Lambda$ vertex, the coupling is 
estimated to be the same as the  N-$\Delta$-$\pi$ 
coupling.  The $\phi$-K-K coupling is also sizeable.
Coupling constants for all reactions are listed in
Table \ref{couplings}.  

The reaction  
$\phi {\rm N} \rightarrow \pi {\rm N} $ is the dominant cross section
in the region $\sqrt{s} > \sqrt{s_0}+1.0$GeV. 
Since reactions
at such high $\sqrt{s}$ do not often occur in the medium, a large 
cross section at extreme $\sqrt{s}$ does not add significantly to 
the collision rate.
The structure emerging in some of the reactions owes to form factors
and dissipative details (imaginary pieces) within the various amplitudes.
Such pieces are necessary to ensure that the interaction range is never
greater than the particle lifetimes allow\cite{rp61}.

\begin{figure}
\centerline{\psfig{figure=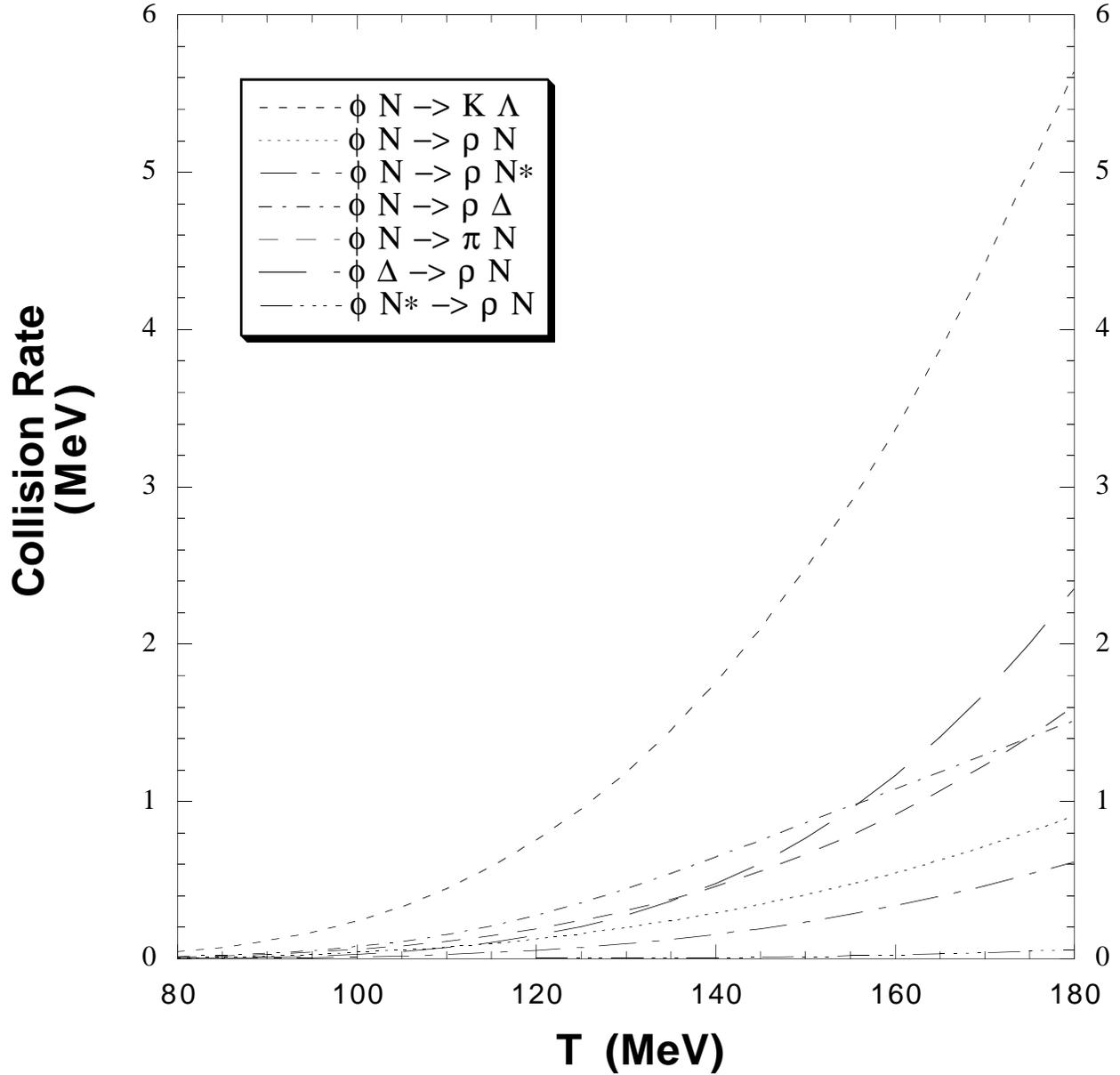}
	    }
\caption{Partial collision rates of the $\phi$ meson with baryons 
in hot,  dense matter.  
This graph assumes a baryon chemical potential of 400MeV. }
\label{thermal400}
\end{figure}

The collision rate for each reaction, as given by \eqn{coll}, is depicted
in Fig.\ \ref{thermal400}.  Again, the largest contribution comes from the
$\phi {\rm N} \rightarrow {\rm K} \Lambda$ reaction.  The population
density  of the various initial particles plays a substantial
role in determining the collision rate.  
The density of $\Delta$ particles is about one half of the nucleon
density at temperature 
$\sim$100 MeV, and rises to equality with the nucleon density
by about 170 MeV.
Thus, while the $\phi \Delta 
\rightarrow \rho$N  cross section is relatively large, its contribution to 
the collision rate does not rise to prominence until high temperature.
Also, the density of the N*(1440) resonance lags behind both the nucleon and
$\Delta$ densities by an order of magnitude.  Thus, the small reaction
cross section for $\phi {\rm N}^* \rightarrow \rho$ N coupled with the
small density of N*'s yields a meager collision broadening effect from this
reaction.

\begin{figure}
\centerline{\psfig{figure=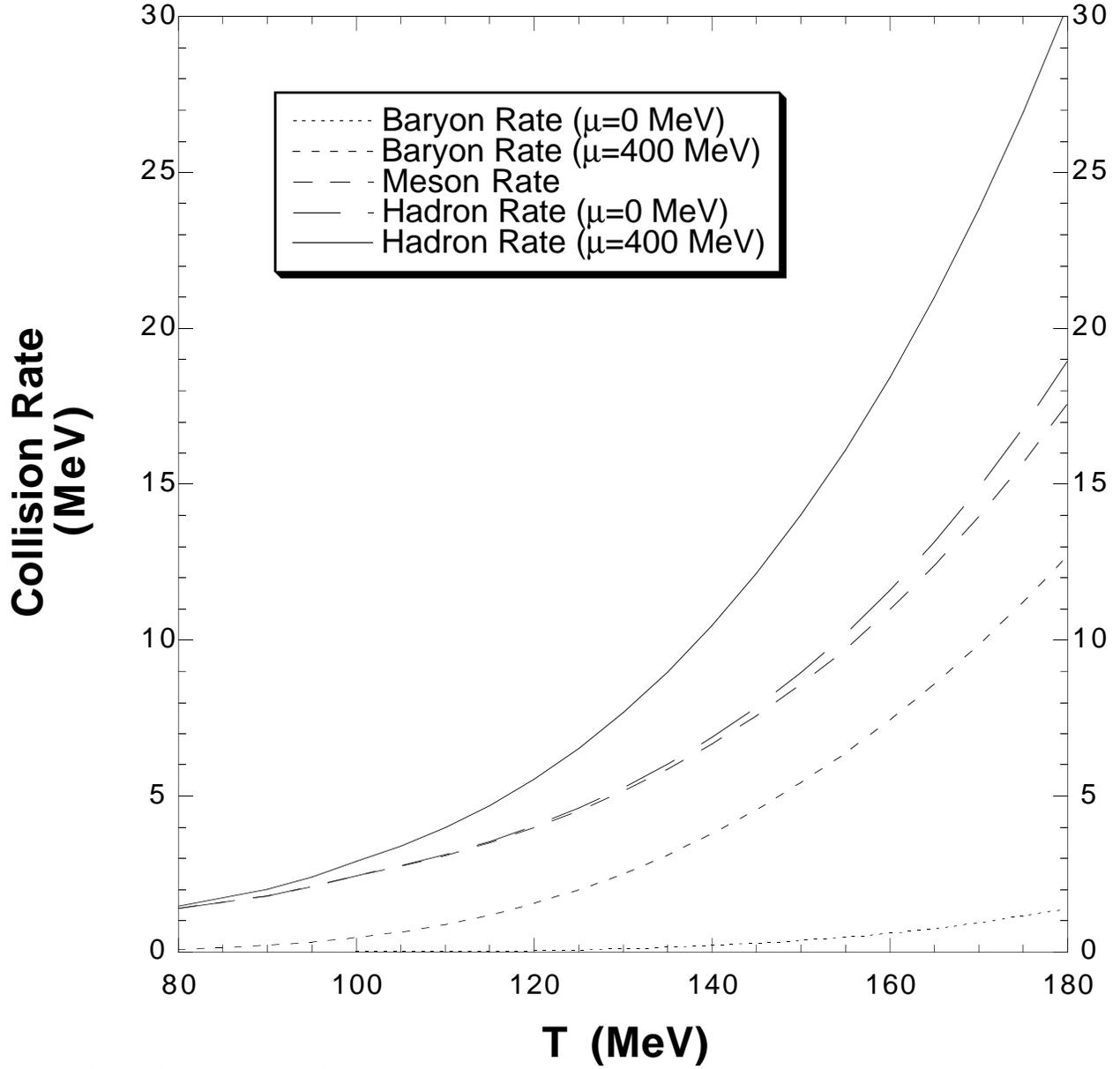}
	    }
\caption{Collision rate of the $\phi$ broken down into the previously
calculated $\phi$ + meson reactions,  
and the current focus, $\phi$ + baryon reactions.}
\label{hadronrate}
\end{figure}

The total collision broadening from all reactions in Fig.\ \ref{thermal400}
is collected and shown in Fig.\ \ref{hadronrate}, along with the previously
calculated collision rate  due to $\phi$ + meson reactions\cite{klh95}.
The free space decay rate of the $\phi$ yields a lifetime
of 45 fm/c.  
Figure \ref{hadronrate}, however, 
depicts a broadening of $\Gamma_\phi$ due to $\phi$ baryon 
interactions by $\simeq~10$ MeV
at a temperature of 170 MeV and a chemical potential of 400 MeV.  Recall, 
\beq
\Gamma^{\rm \footnotesize{total}}=
\Gamma^{\rm \footnotesize{decay}}+\Gamma^{\rm \footnotesize{collision}}
\eeq
where $\Gamma_\phi^{\rm \footnotesize decay} = 4.43$ MeV.
Then the effective width has broadened to $\sim $14 MeV.
Thus, due to baryon collisions alone there is possibly enough evidence 
to suggest that $\phi$ becomes kinetically thermal.

\begin{figure}
\centerline{\psfig{figure=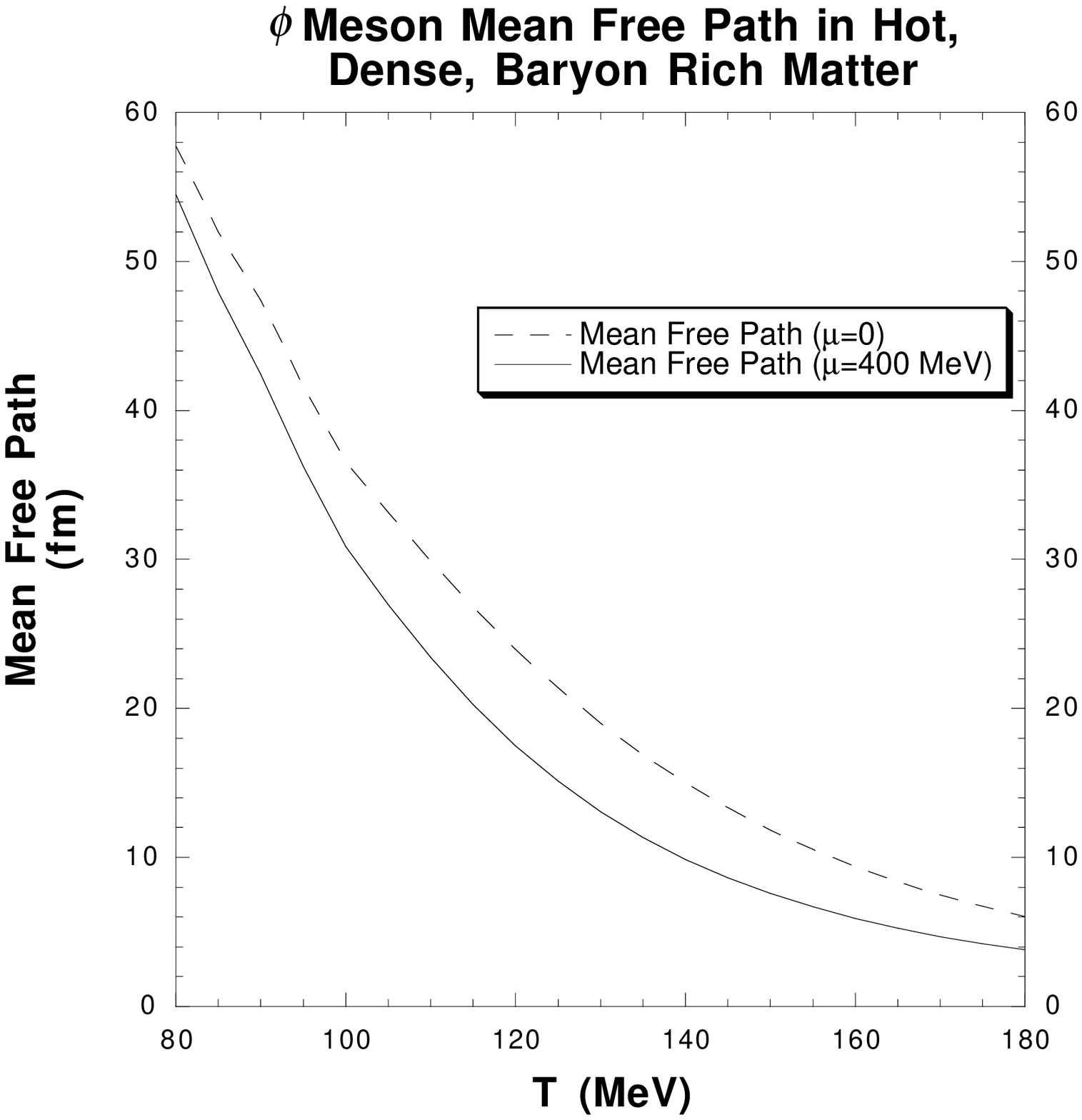}
	    }
\caption{The mean free path of the $\phi$ in hot,  dense, baryon rich
matter.  At high temperatures this provides a good indication that the
$\phi$ partially thermalizes.}
\label{meanfree}
\end{figure}

The collision broadening due to meson interactions has been computed 
previously\cite{klh95}.  The estimate for $\phi$ + meson reactions
is plotted along with the current $\phi$ + baryon prediction in 
Fig.\ \ref{hadronrate}.  Through collisions with hadrons the width
of the $\phi$ is broadened by $\simeq$ 24 MeV at 170 MeV with a baryon
chemical potential of 400 MeV.
One finds a 24 MeV broad $\phi$ distribution and concludes
that there is  compelling evidence to suggest that the $\phi$ meson  
can fully thermalize even given the brief existence of the 
reaction heavy ion systems.
Employing the estimate of \eqn{mfp}, the mean free path of the $\phi$
is plotted in Fig.\ \ref{meanfree}.  At 170 MeV temperature and a
chemical potential of 400 MeV, these results indicate that the $\phi$ has a
mean free path of a mere $\simeq 5$ fm!  At the same temperature, and with 
zero chemical potential, the mean free path rises to $\simeq 7$ fm.
Again, this presents fairly strong evidence for kinetic thermalization
of the $\phi$.

\section{Summary}

We have used  effective 
Lagrangian methods within kinetic theory to describe dynamics of the 
$\phi$ meson in a hot and dense, baryon rich nuclear medium.
We have shown that the 4.43 MeV free space width of the $\phi$
is broadened by about 10 MeV through collisions with baryons
for conditions typical of high energy heavy ion collisions.  These results,
coupled with previous results for meson reactions\cite{klh95}, indicate
a broadening of the $\phi$ width by about 24 MeV in the same conditions.  
Thermalization for the $\phi$ meson is a reasonable conclusion.



%
%
\begin{table}
\caption{Coupling constants.}
\label{couplings}
\begin{tabular}{cc}
Coupling Constant & Numerical Value \\ \tableline
${g^2_{N\pi\Delta}}/{4\pi}$ & 15 \\
${g^2_{\phi\rho\pi}}/{4\pi}$ & 5.6 $\times 10^{-3}$ \\
${g^2_{\phi KK}}/{4\pi}$ & 2.71 \\
${g^2_{NN\rho}}/{4\pi}$  (with ${f_{NN\rho}}/g_{NN\rho}$ = 5.0)
& 0.85 \\	
${g^2_{NN\pi}}/{4\pi}$ & 14.4 \\
${g^2_{N^*N\pi}}/{4\pi}$ & 13 \\
${g^2_{NK\Lambda}}/{4\pi}$ & 15 \\
\end{tabular}
\end{table}


\end{document}